\documentclass[]{spie}  

\usepackage{amsmath,amsfonts,amssymb}
\usepackage{graphicx}
\usepackage[colorlinks=true, allcolors=blue]{hyperref}

\title{High Contrast Imaging at the Photon Noise Limit with WFS-based PSF Calibration}

\author[a,b,c,d]{Olivier Guyon}
\author[e]{Barnaby Norris}
\author[e]{Marc-Antoine Martinod}
\author[a]{Kyohoon Ahn}
\author[a]{Vincent Deo}
\author[a,f,g]{Nour Skaf}
\author[a]{Julien Lozi}
\author[a,d,f]{S\'ebastien Vievard}
\author[b]{Sebastiaan Y. Haffert}
\author[a]{Thayne Currie}
\author[b]{Jared R. Males}
\author[e]{Alison Wong}
\author[e]{Peter Tuthill}

\affil[a]{Subaru Telescope, National Astronomical Observatory of Japan, National Institutes of Natural Sciences (NINS), 650 North A`oh\={o}k\={u} Place, Hilo, HI 96720, United States}
\affil[b]{Steward Observatory, University of Arizona, Tucson, AZ 87521, United States}
\affil[c]{College of Optical Sciences, University of Arizona, Tucson, AZ 87521, United States}
\affil[d]{Astrobiology Center of NINS, 2 Chome-21-1, Osawa, Mitaka, Tokyo, 181-8588, Japan}
\affil[e]{Sydney Institute for Astronomy, Institute for Photonics and Optical Science, School of Physics, University of Sydney, NSW 2006, Australia}
\affil[f]{Observatoire de Paris, LESIA, 5 Place Jules Janssen, 92190 Meudon, France}
\affil[g]{Department of Physics and Astronomy, University College London, London, United Kingdom}

\authorinfo{Further author information: (Send correspondence to O.G.)\\O.G.: E-mail: guyon@naoj.org}

\begin{document} 
\maketitle

\begin{abstract}
Speckle Noise is the dominant source of error in high contrast imaging with adaptive optics system. We discuss the potential for wavefront sensing telemetry to calibrate speckle noise with sufficient precision and accuracy so that it can be removed in post-processing of science images acquired by high contrast imaging instruments. In such a self-calibrating system, exoplanet detection would be limited by photon noise and be significantly more robust and efficient than in current systems. We show initial laboratory and on-sky tests, demonstrating over short timescale that residual speckle noise is indeed calibrated to an accuracy exceeding readout and photon noise in the high contrast region. We discuss immplications for the design of space and ground high-contrast imaging systems.
\end{abstract}


\section{Scientific motivation}
\label{sec:fundlimits}

Direct imaging of exoplanets and circumstellar dust with space and ground telescopes is challenging due the high flux ration between the central star and the object of interest (planet or disk), combined with the small angular separation between the sources. High contrast imaging (HCI) systems are designed to overcome these challenges by combining optical starlight suppression with wavefront control. At optical and near-IR wavelengths, detection with HCI systems is almost always limited by speckle noise: residual uncontrolled and uncalibrated starlight behind the coronagraph mask imposes a contrast floor below which detections become too unreliable. 

Differential detection techniques can be employed to mitigate speckle noise. One approach is to use either high spectral reolution signatures in the exoplanet light \cite{Wang_2017} or polarized light in starlight by circumstellar dust or exoplanet atmophseres \cite{2020AA...634A..69H}. The fraction of an exoplanet light containing polarization of spectral signatures is small, typically up to a few percent, so these differential imaging techniques, although generally quite reliable, are inefficient and poorly suited to detect the faintest sources. Another approach is to separate the speckles from real sources by angular differential imaging (ADI) \cite{2006ApJ...641..556M}, which detects the planet image thanks to the known field rotation, or spectral differential imaging (SDI) \cite{1999PASP..111..587R}, which identifies speckles through their wavelength geometrical scaling. The differential geometrical effects (rotation or scaling) that ADI and SDI rely on are proportional to angular separation, so they do not provide significant contrast gains at the smallest angular separations where some of the most valuable exoplanets lie. The underlying assumptions of PSF stationarity or wavevength-scaling are approximations, so ADI and SDI also do not perfectly remove speckle noise.
A third line of research is to use the mutual coherence between starlight and speckles to identify them, by spatial \cite{2006dies.conf..553B} or temporal \cite{2004ApJ...615..562G} modulation. These promising approaches have recently seen significant progress, but remain challenging to implement on-sky, as speckles can be mis-identified as incoherent due to chromaticity and finite temporal resolution.

In this paper, we explore a complementary approach, where wavefront sensing telemetry could be used to reconstruct the speckle cloud with sufficient accuracy and precision so that it can be numerically removed down to the photon noise residual. If successful, this would allow reliable detection of Earth-like habitable exoplanets orbiting sun-like stars with a 6-m space telescope without relying on extreme long-term stability of the telescope optics, or orbiting M-type stars with a ground-based 30-m aperture equiped with extreme adaptive optics.

To illustrate the scientific potential of high contrast imaging at the photon noise limit, we consider examples representative of space and ground telescopes imaging habitable exoplanets:
\begin{itemize}
    \item An Earth-like planet orbiting a Sun-like star at 10pc distance, observed by a 6-m space telescope
    \item An Earth-size planet orbiting in the habitable zone of a M4 type star at a 4pc distance, observed by a 30-m ground-based telescope
\end{itemize}

\begin{table}[ht]
\caption{Observation examples} 
\label{tab:HCIobs}
\begin{center}       
\begin{tabular}{|l|c|c|} 
\hline
            & Space-6m-Earth-G2 &  Ground-30m-Earth-M4\\
\hline
\hline
Star &  G2 at 10pc &  M4 at 4pc \\
\hline
Bolometric luminosity [$L_{Sun}$] &  1.000 & 0.0072  \\
\hline
Planet orbital radius [au] &  1.0 &  0.085 \\
\hline
Maximum angular separation [arcsec] &  0.1 & 0.021 \\
\hline
Reflected light planet/star contrast &  1.5e-10 & 2.1e-8 \\
\hline
\hline
Telescope diameter [m] & 6 & 30 \\
\hline
Science spectral bandwidth &  20\% &  20\%  \\
\hline
Central Wavelength &  797 nm (I band) &  1630 nm (H band) \\
\hline
Maximum angular separation [$\lambda$/D] & 4.5 & 1.9 \\
\hline
Efficiency &  20 \% &  20 \%  \\
\hline
Total Exposure time &  10 ksec &  10 ksec  \\
\hline
\hline
Star brightness & $m_I = 4.04$ & $m_H = 5.65$ \\
\hline
Photon flux in science band (star) & 1.06e9 ph/s & 5.62e9 ph/s \\
\hline
Photon flux in science band (planet) & 0.16 ph/s & 118 ph/s \\
\hline
Background zodi+exozodi [contrast] & 3.1e-10 & \\
Background starlight [contrast] & 1e-10 & 1e-5\\
Total background         & 4.1e-10 & 1e-5\\
\hline
Background flux in science band & 0.43 ph/s & 56200 ph/s\\
\hline
\hline
{\bf Photon-noise limited SNR (10 ksec)} & 20.8 & 49.7 \\
\hline
{\bf Exposure time for photon-noise limited SNR=10} & 38 mn & 7 mn\\
\hline
{\bf Required PSF calibration accuracy (SNR=10)} & 0.15 & 5e-3\\
\hline
\end{tabular}
\end{center}
\end{table}

Photon-noise limited sensitivity derivations are shown in table \ref{tab:HCIobs}.


For the space-based observation, the background flux is dominated by the combined zodiacal and exozodiacal light components, which contribute 2.7 $\times$ more light that the planet. A SNR $=$ 10 detection requires 38 mn integration.
The ground-based telescope benefits from a larger collecting area, so the planet photon rate is much larger ($>$100 ph/s), but the background is also significantly higher due to residual atmospheric turbulence, assumed here to be at the $1e5$ contrast level. A 7 mn integration allows for SNR=10 detection of the planet.

In order to calibrate speckle noise to 10 $\times$ below the planet flux, the residual PSF light must be calibrated to 15\% accuracy for the space observation, and to 5e-3 accuracy for the ground observation. We discuss in this paper steps towards achieving this accuracy, which is especially challenging for the ground observation example discussed in this section.

\section{Wavefront Sensor to High Contrast Image Mapping}

\subsection{Self-Calibration Approach}

\begin{figure}[ht]
    \centering
    \includegraphics[width=17cm]{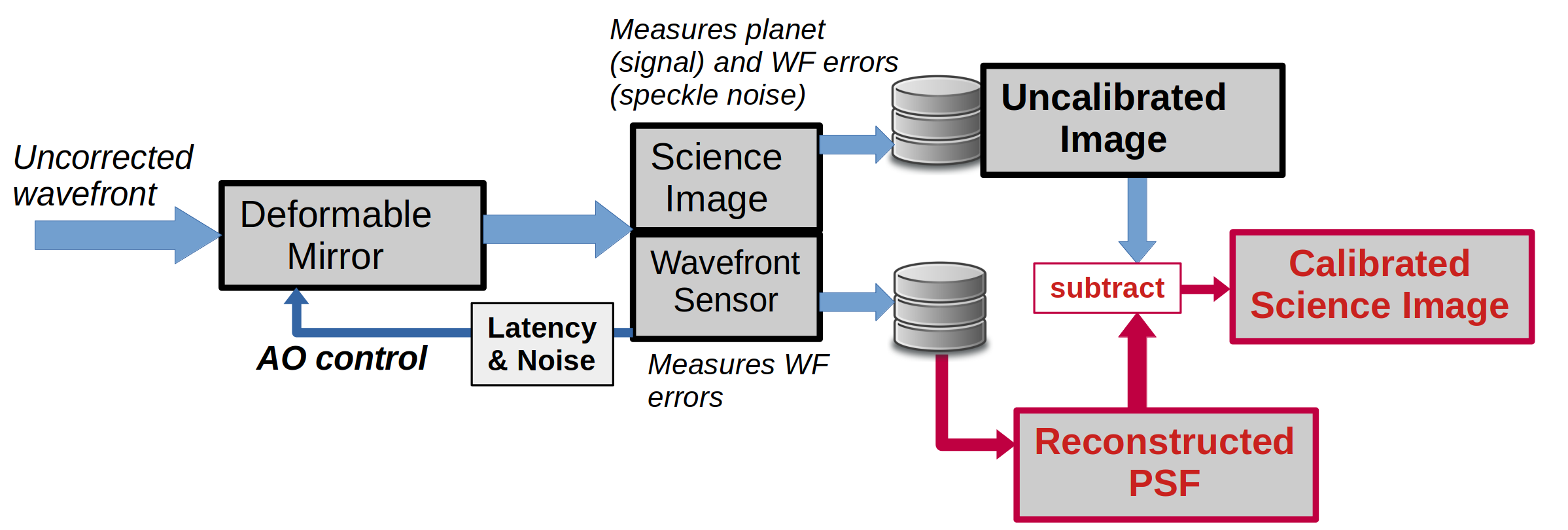}
    \vspace*{0.3cm}
    \caption{WFS-based PSF calibration principle. In a conventional high contrast imaging system, wavefront sensor (WFS) measurements are used to drive a wavefront control loop. We discuss in this paper how the WFS telemetry can also be used to reconstruct the point spread function (PSF) so that speckle noise can be subtracted in high contrast images (red boxes).}
    \label{fig:calibprinciple}
\end{figure}

High-contrast imaging (HCI) systems rely on wavefront sensor(s) (WFS) to measure optical aberrations. The measurements serve as the input to a wavefront control loop maintaining near-flat wavefront. WFS measurements usually serve no other purpose. Our calibration approach is to use the WFS telemetry to reconstruct the point spread function (PSF) with sufficient accuracy and precision to subtract residual starlight. These additional steps are shown in Figure \ref{fig:calibprinciple}, colored in red.

The main challenge to this approach is that it reaquires accurate knowledge of the relationship between WFS telemetry and PSF. The relationship must be inferred either by modeling, or learned by analysis of telemetry. We explore in this paper the latter option.

The WFS-to-PSF relationship is non-linear, so it is not possible to reconstruct the average PSF from an average of the WFS telemetry. For example, a rapid zero-mean oscillation of a pupil plane sine wave mode will generate a pair of focal plane speckles, even through the average wavefront state is free of aberration. This non-linearity requires the WFS-to-PSF reconstruction to be performed sufficiently fast to capture wavefront temporal variations. For a ground-based system, the WFS-to-PSF reconstruction should be performed at $\approx$ kHz speed, while it can be considerably slower on a space-based system. Reconstructed intensity PSFs may then summed to temporally match the science integration and yield a reconstructed PSF that can be subtracted from the science data. The resulting calibrated science image should then be free of speckle noise. In this scheme, detection contrat is limited by the photon noise of the speckle field, as opposed to the speckle field itself.

We note that previous PSF reconstruction efforts have followed a different approach, where the long exposure PSF is estimated from the temporal variance of each wavefront mode \cite{2006AA...457..359G}. This approach does not require each WFS frame to be processed, but requires an accurate model of the WFS-to-PSF relationship, and makes several approximations that do not allow for high fidelity PSF reconstruction for high contrast imaging, namely: WFS linearity, quadratic relationship between WF and PSF intensity, and statistical independence between modes. This is not the approach explored in this paper.


\subsection{Post-processing vs. closed loop control}

Since, as shown in Figrue \ref{fig:calibprinciple}, the WFS telemetry serves as the input to the wavefront control (WFC) loop, we discuss here what is the value of also using WFS telemetry for PSF reconstruction. We must consider what information is contained in WFS telemetry that is valuable for PSF reconstruction but cannot be used for WFC.

There are two fundamental advantages in using WFS telemetry for PSF reconstruction over WFC :
\begin{itemize}
    \item {{\bf Correction Null Space.} WFS telemetry contains information about the WF state that is not correctable by the WFC system. For example, WFS frames may extend to higher spatial frequencies than correctable by the deformable mirror(s). WFS frames may also measure amplitude errors. Such errors are in the correction null space, but affect the PSF.}
    \item {{\bf Immunity to Latency and Arrow of Time for WF estimation} In a WFC loop, each DM command is based on past WFS measurements. Due to hardware latencies such as camera exposure time, the current WFS measurement is not accessible (it will only be available at a later time). Post-processing PSF reconstruction is immune from such latency.}
\end{itemize}

\begin{figure}[ht]
    \centering
    \includegraphics[width=17cm]{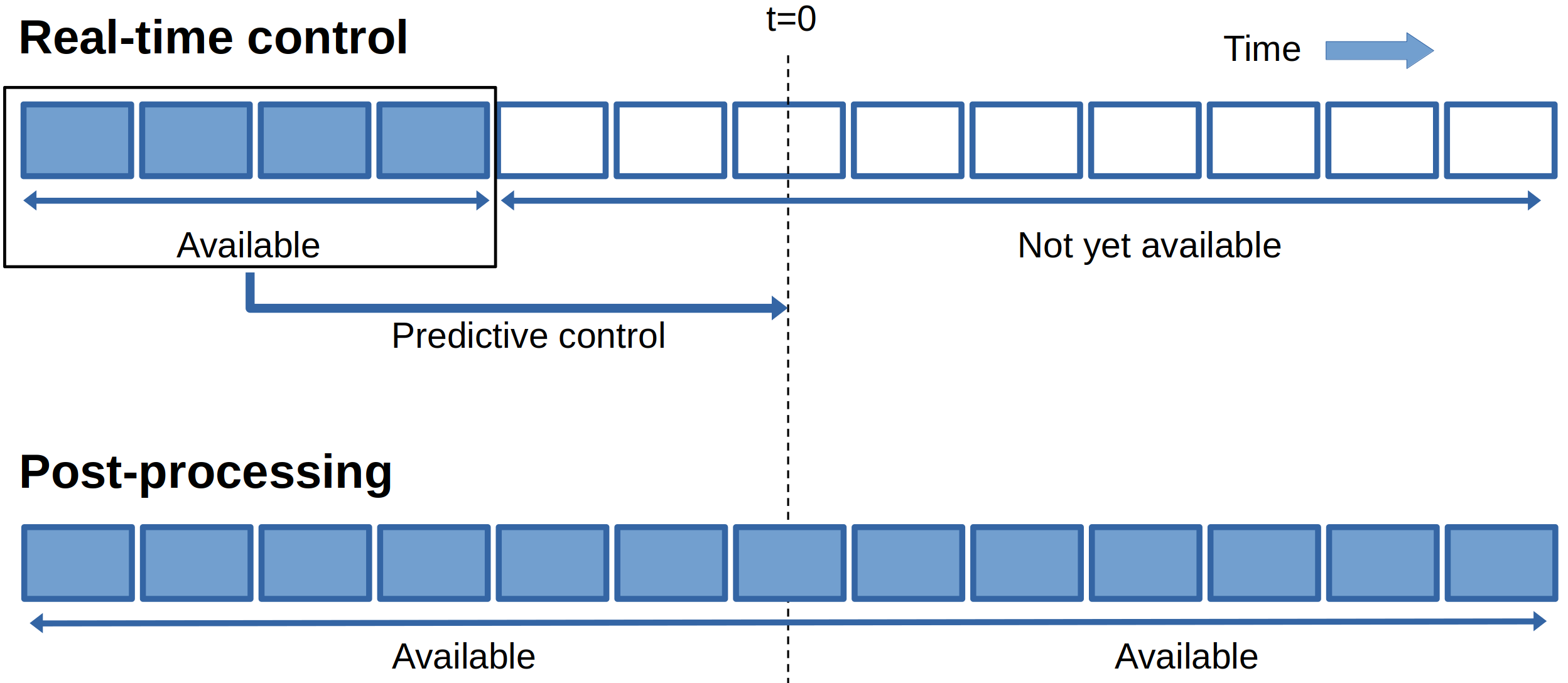}
    \vspace*{0.3cm}
    \caption{Use of WFS temetry for control vs. PSF calibration. In wavefront conrol (top), only past measurements are available to estimate the current WF state. Due to latency, the current and most recent measurements are not yet available. Post-processing (bottom) does not suffer from this limitation: past, current and future measurements are available for PSF estimation.}
    \label{fig:arrowoftime}
\end{figure}

The arrow of time concept is illustated in Figure \ref{fig:arrowoftime}. In a WFC correction loop (top), the only WFS measurements available to the control algorithm are in the past, and the WFS telemetry corresponding to the current (and recent) time is not yet available. The control command relies on old and noisy estimates, yielding relatively large WF esimation errors. While predictive control algorithms may help mitigate this limitation, the correction is fundamentally blind to recent and current WF changes. In post-processing (bottom), past, current and future measurements are all available, allowing for a more accurate WF estimation free of latency. By interpolating/averaging between all meassurements, WF estimates can also average down measurement noise.

In addition to the null space and arrow of time advantages, post-processing has access to a large sample set of WFS-to-PSF samples, so an accurate model of the WFS-to-PSF relationship can be infered from the telemetry.

\section{On-Sky Experimental Validation of WFS-to-PSF mapping}

\subsection{Solution Uniqueness}
\label{ssec:mappinguniqueness}

The WFS-to-PSF calibration approach requires a unique relationship from WFS measurement to PSF intensity. If such a relationship exists, and is stable over time, it can be learned from synchronized samples of WFS and PSF realizations. The first step in developing a PSF calibration process is therefore to test is the relationship exists: does WFS telemetry impose a unique PSF solution ?

\begin{figure}[ht]
    \centering
    \includegraphics[width=17cm]{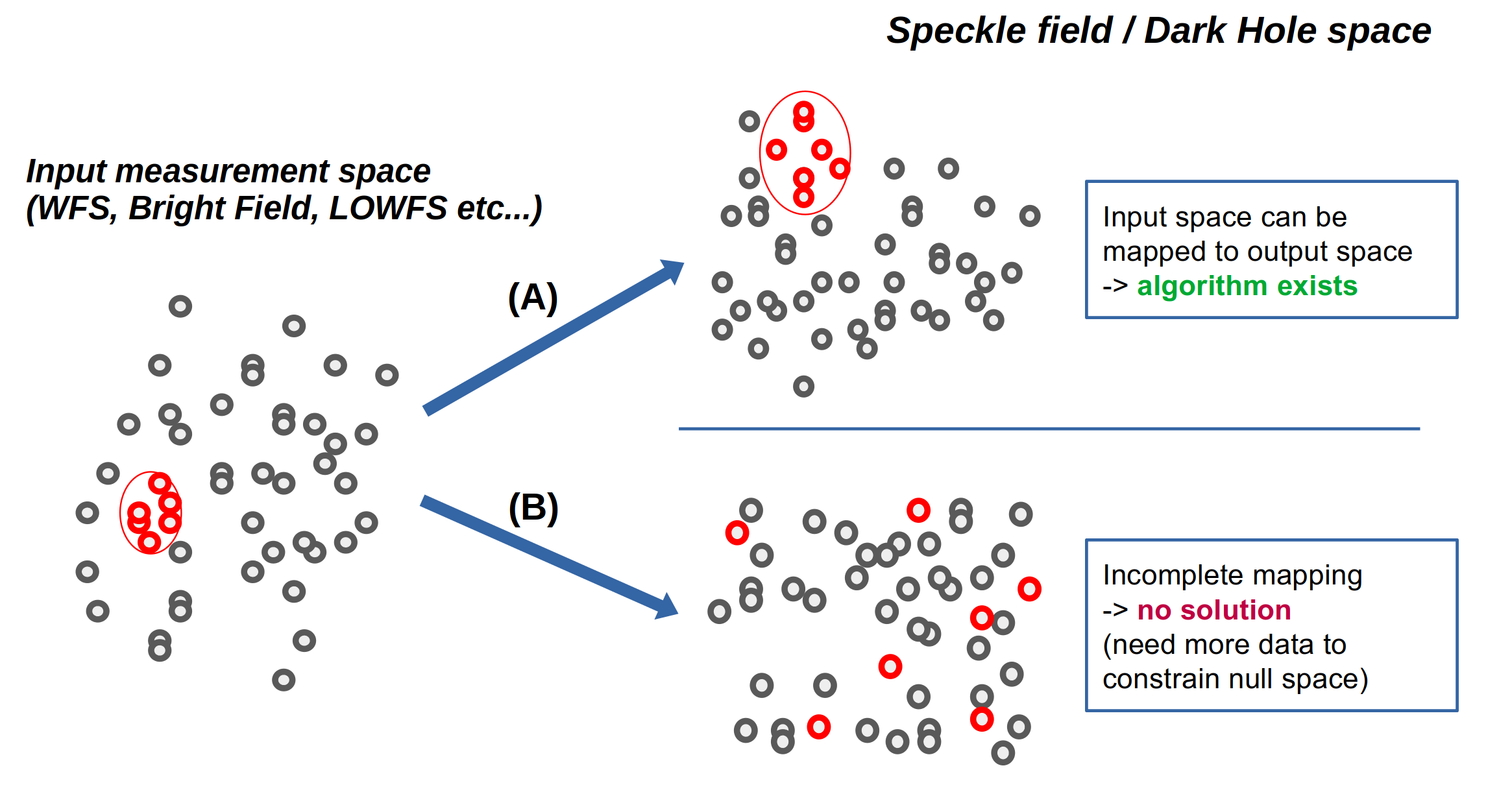}
    \vspace*{0.3cm}
    \caption{Solution uniqueness. In case (A) there is a unique mapping from WFS to PSF, while in case (B) the mapping is not unique. }
    \label{fig:uniqueness}
\end{figure}

Figure \ref{fig:uniqueness} illustrates two possible cases, and indicates how statistical tests can be engineered to test solution uniqueness. WFS measurements are shown on the left, and corresponding PSF intensity on the right. In case (A), there is a unique mapping between input and output spaces and WFS-to-PSF calibration is possible. In case (B) there is no such mapping, and WFS telemetry is insufficient to constrain the PSF. If a unique mapping exists, clusters of nearby points in the input space map to clusters on nearby points in the output space. If no such mapping exists, nearby points in the input space can map to distant points in the output space.

A statistical test to validate the existence of a WFS-to-PSF calibration algorithm is to identify nearby WFS realizations and check that the corresponding output PSFs are similar. We note that this empirical test is not explicitely relying on a WFS-to-PSF reconstruction algorithm. Yet, a multidimentional mapping could be constructed by grouping input points in small clusters and measuring the corresponding output state. Each new measurement would then be matched to one of the input clusters and the corresponding solution would be read from a lookup table. In a high dimension space, this algorithms is not practically feasible and some interpolation is required.

\subsection{Experimental Validation of Unique WFS-to-PSF Mapping}

The first step toward an on-sky WFS-to-PSF calibration algorithm is to perform the statistical mapping test described in \S \ref{ssec:mappinguniqueness}. We used the Subaru Coronagraphic Extreme Adaptive Optics (SCExAO) system \cite{2018SPIE10703E..59L} on the Subaru Telescope for this test. The input is a pyramid WFS operating in visible light, and the output is a focal plane image at 750nm wavelength acquired with the VAMPIRES instrument \cite{2020SPIE11203E..0SN}. The test was performed in support of the PSF-sharpening iterative DrWHO algorithm where the AO control loop rewards WFS frames corresponding to sharp PSF realizations \cite{2022AA...659A.170S, 10.1117/12.2595008}.

\begin{figure}[ht]
    \centering
    \includegraphics[width=17cm]{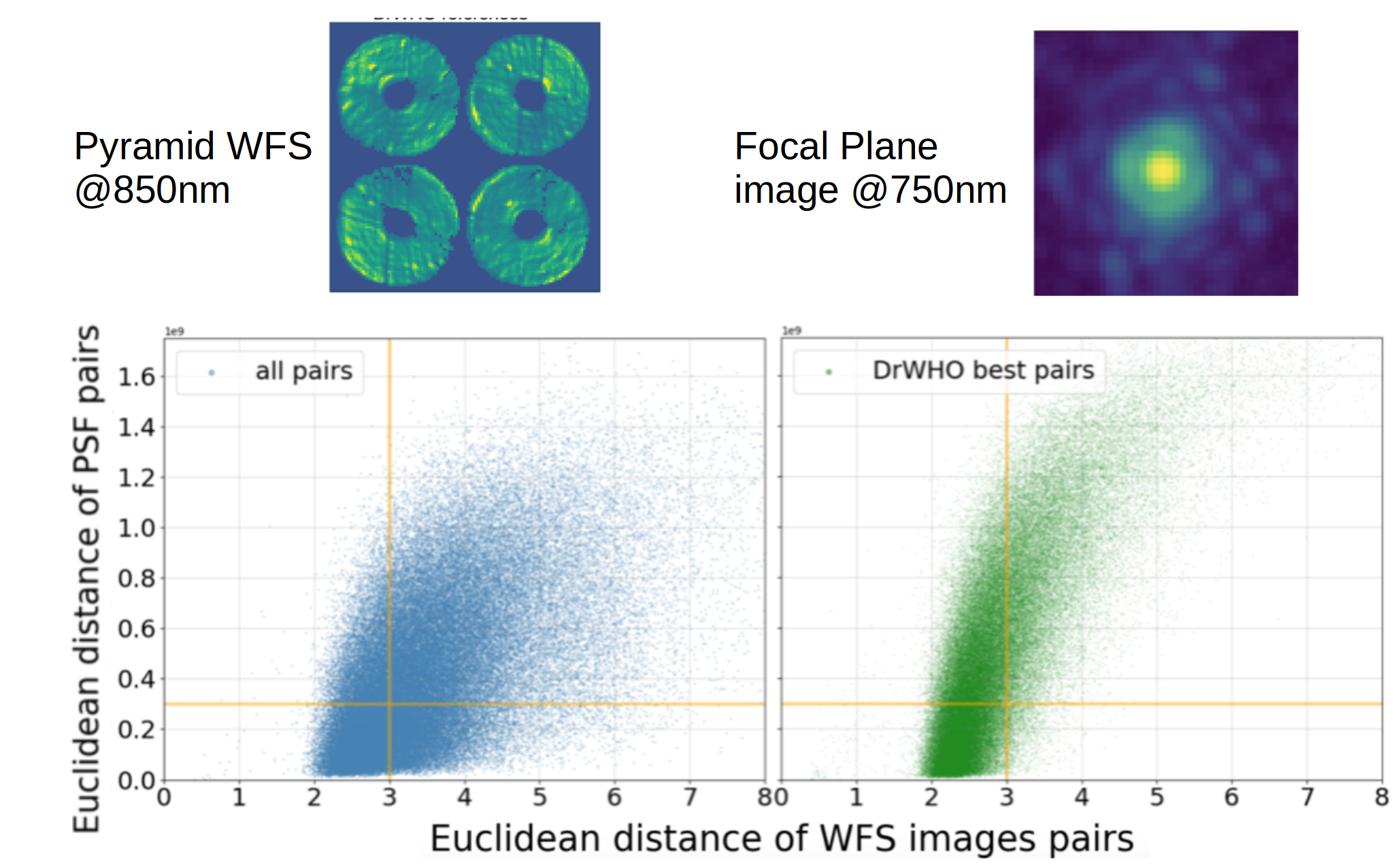}
    \vspace*{0.3cm}
    \caption{Solution uniqueness between WFS and PSF: on-sky validation using SCExAO's pyramid WFS (top right) and VAMPIRES's visible light imaging (top right). Adapted from the DrWHO algorithm on-sky validation \cite{2022AA...659A.170S}.}
    \label{fig:DrWHO}
\end{figure}

Results are shown in Figure \ref{fig:DrWHO}. Example WFS and PSF frames are shown at the top. We first constructed synchronized (WFS, PSF) realizations, and then compared pairs of such samples. The bottom left plot shows how euclidian distance between WFS frames compares to euclidian distance between the corresponding PSF frames. The left side of the diagram corresponds to pairs with similar WFS realizations, while the bottom part of the diagram shows similar PSFs. The distribution of points reveals that WFS-similarity implies PSF-similarity: all points to the left side of the cloud exhibit small PSF distance. The converse does not hold: PSF-similarity (bottom of the cloud) does not correspond to WFS-similarity. There are numerous points in the lower right part of the cloud. This experimental results confirms that there is a unique mapping from WFS to PSF, but also reveals there is not unique mapping from PSF to WFS.

This result is consistent with expectations: the WFS provides an unambiguous measurement of WF state for the modes that are sensed. The focal plane image does not map unambiguously to WF state: for example, the signature of the focus WF mode on an otherwise perfect PSF is sign-ambiguous, as the opposite focus values yield the same image. In this experiment, the DrWHO algorithm selected high quality PSF realizations for which the focus WF mode was close to zero. In this second sample (bottom right diagram), WFS similarity is enforced by PSF similarity.

\subsection{PSF Reconstruction Validation in Standard Imaging Mode}

\label{sec:WFStoPSF}

\begin{figure}[ht]
    \centering
    \includegraphics[width=10cm]{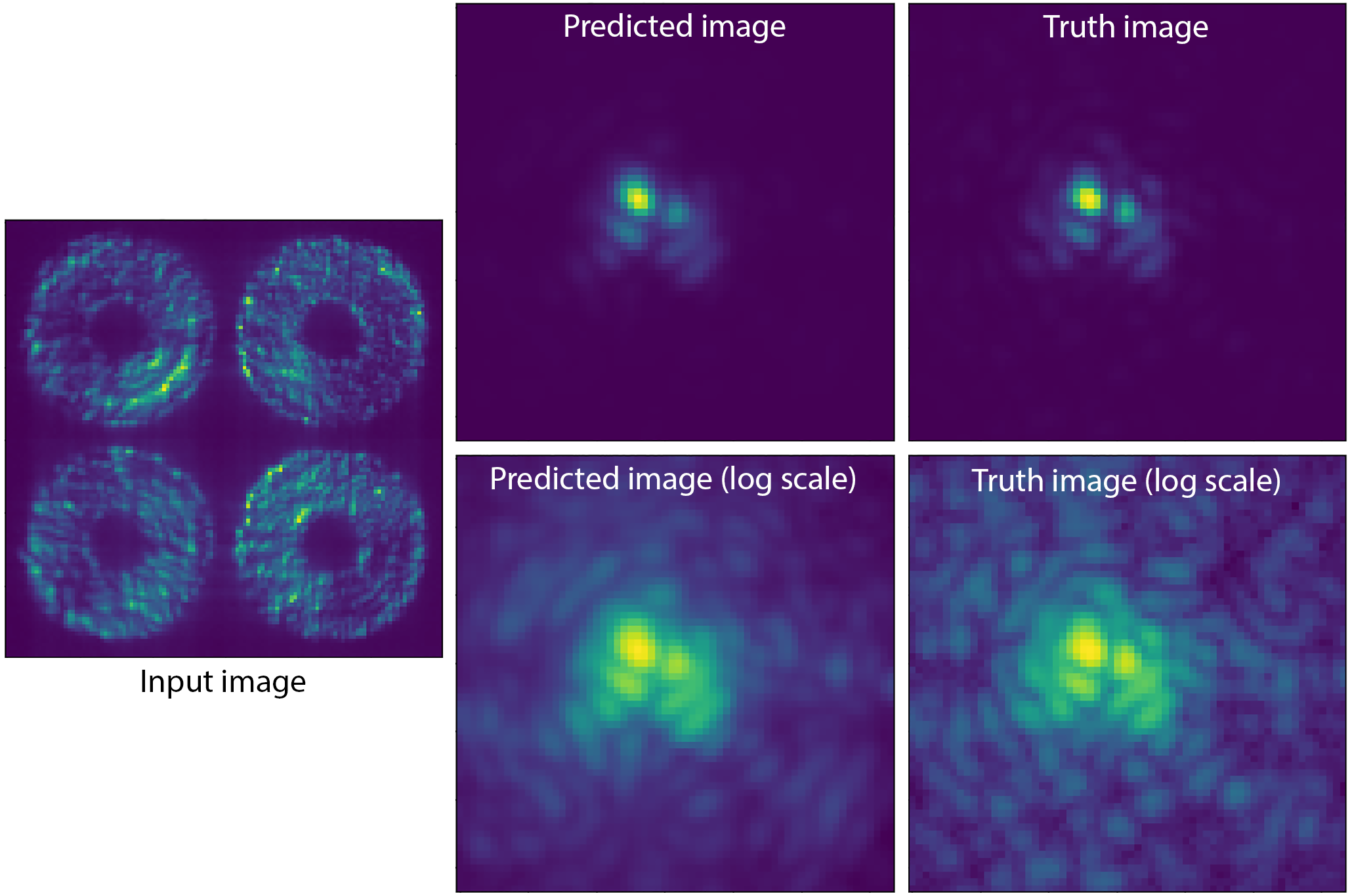}
    \vspace*{0.3cm}
    \caption{Prediction of the PSF from the Pyramid WFS data for an on-sky observation, using a neural network. The predicted PSF image (centre) is determined entirely from the current WFS image (left), and is seen to closely match the true PSF measured at that instant (right column). This example shows a PSF with a large amount of wavefront error (including strong coma) to provide a clear illustration.}
    \label{fig:wfspsf}
\end{figure}

Having verified that a unique mapping exists from WFS telemetry to PSF, we can attempt to reconstruct the PSF from WFS telemetry. Results of a PSF reconstruction for on-sky data are shown in Figure \ref{fig:wfspsf}. Here, a fully-connected neural network consisting of two 2000-unit layers was trained on 5 minutes of on-sky data, consisting of synchronised images from the pyramid wavefront sensor camera and VAMPIRES visible camera\cite{2015MNRAS.447.2894N} (wavelength 750~nm), running at approximately 500 Hz frame rate. The network used ReLU activation functions and dropout between each layer as a regularizer, the latter proving to be crucial for successful reconstruction. While the fully-connected network shown here provides good results, certain advantages (such as reduced parameter number and resistance to pupil alignment drift) could be expected from a convolutional neural network, which is the focus of a current study.

The main features of the PSF speckle cloud are recovered by the reconstruction. The PSF core shape and bright diffraction features are accurately reproduced, but fainter speckles are less accurately reconstructed.

\section{High Contrast Validation}
\label{sec:specklecalib}

\subsection{Self-Calibration in a Coronagraphic System}

We extend here the WFS-to-PSF calibration to coronagraph systems optimized for high contrast imaging. When considering HCI systems, we refer to the input WFS as the bright field (BF) and the output speckle cloud as the drak field (DF). The self-calibration's goal is to derive the DH from BF. An ideal system for self-calibration would have a stable BF-to-DH relationship. This can be challenging with a configuration such as the one described in \S \ref{sec:WFStoPSF}, where the WFS and PSF use different optical trains and cameras.

We therefore used a configuration where BF and DF are co-located on the same detector and share the same optical train. The SCExAO near-IR Lyot coronagraph was configured with speckle control to produce a deep contrast area (DF) over one side of the focal plane mask, with the opposite side of the focal plane image remaining relatively bright and serving as the input BF, as shown in Figure \ref{fig:DHcalibLDFC} top left image. The experiment is closely related to linear dark field control (LDFC) \cite{2021AA...646A.145M}, where a linear control loop uses the BF as input for wavefront control. LDFC has been demonstrate to stabilize the DH both in the laboratory \cite{2020PASP..132j4502C} and on-sky \cite{2021AA...653A..42B}. Here, we explore extending LDFC as a DH calibration algorithm, without assuming linearity.

\subsection{Laboratory Validation with a Lyot Coronagraph}

\begin{figure}[ht]
    \centering
    \includegraphics[width=16cm]{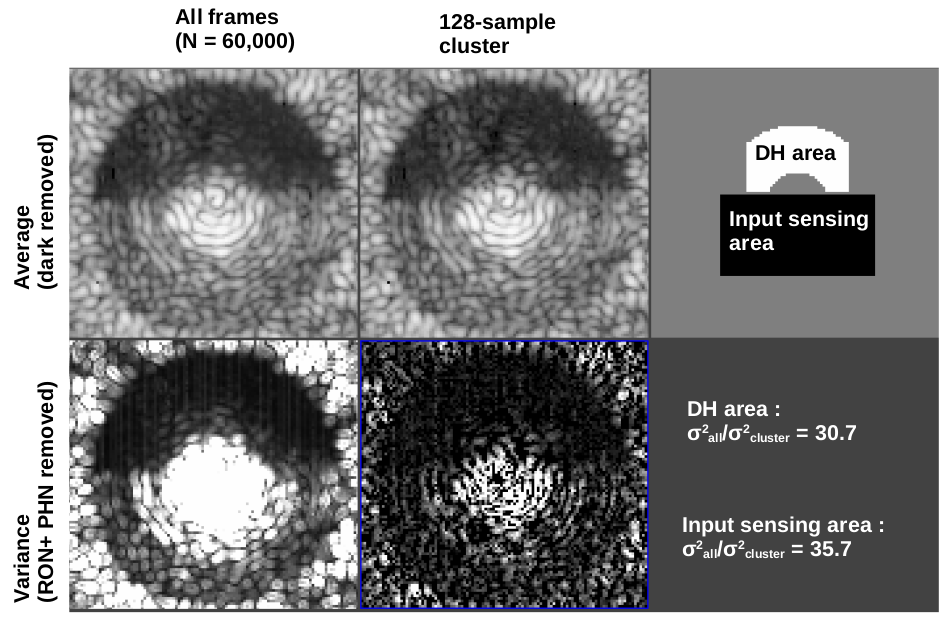}
    \vspace*{0.3cm}
    \caption{Self-calibration of high contrast images in a half dark hole configuration on the SCExAO bench. A near-IR Lyot type coronagraph is used to block starlight. Focal plane wavefront control creates a deep contrast area above the coronagraph mask, and a relatively bright area below it. The DH is calibrated from the input sensing area (BF) below the optical axis. A clustering algorithm is employed to derive BF-to-DH mapping.}
    \label{fig:DHcalibLDFC}
\end{figure}

Results are compiled in Figure \ref{fig:DHcalibLDFC}. The average of all 60000 frames (top left) shows the BF in the lower half of the image and the DH in the top half. The intensity variance across the 60000 images is shown for each pixel of the image at the bottom left, with both readout noise and photon noise variance terms subtracted to reveal actual speckle intensity variance.

A clustering algorithm is used to derive the mapping between BF and DH, as detailed in a previous publication \cite{2021SPIE11823E..18G}. Results are shown in Figure \ref{fig:DHcalibLDFC}, comparing the full dataset (left column) with one of the clusters serving as an entry in the BF-to-DH mapping table (center column). The cluster consists of 128 samples with similar BF realizations: the variance within the BF is 35.7 $\times$ smaller within the cluster than across the full dataset. The corresponding measured DH variance is 30.7 $\times$ smaller within the cluster set than across the full input dataset. This last result demonstrates that images with similar BFs also have similar DHs. {\bf Image selection using BF intensity successfully constrains DH intensity, demonstrating that a BF-to-DH calibration algorithm can be derived to calibrate residual speckles in high contrast images}.

\subsection{Null Calibration with Photonic Nulling Interferometer: On-Sky Validation}
\label{sec:GLINT}

A photonic nuller is an alternative solution to the high contrast imaging challenge. Unlike a coronagraph constructed from bulk optics between which light freely propagates, the photonic nuller couples starlight into a small number of coherent singlemode waveguides. The waveguides are coherently combined to produce starlight destructive interference in null output(s). Bright starlight is directed to bright outputs which measure the intensity in input waveguides (photometry output(s)) and phase offset between input waveguides (WFS output(s)). The photonic nuller concept and its implementation are discussed in publications from the GLINT instrument team\cite{2020MNRAS.491.4180N, Martinod2021NatCo}.

The photonic nuller is well-suited for self calibration :
\begin{itemize}
    \item Starlight is coupled in a small number of coherent waveguides. At each wavelength, light into the photonic device input is fully described by phase and amplitude (and possibly polarization), so the number of dimension in the input is a few times the number of waveguides
    \item The relationship between input variables (phase and amplitude of each waveguide) and output intensities is entirely established within the photonic chip so it is significantly more stable than an optical train of optical components subject to relative misalignments.
\end{itemize}

\begin{figure}[ht]
    \centering
    \includegraphics[width=8cm]{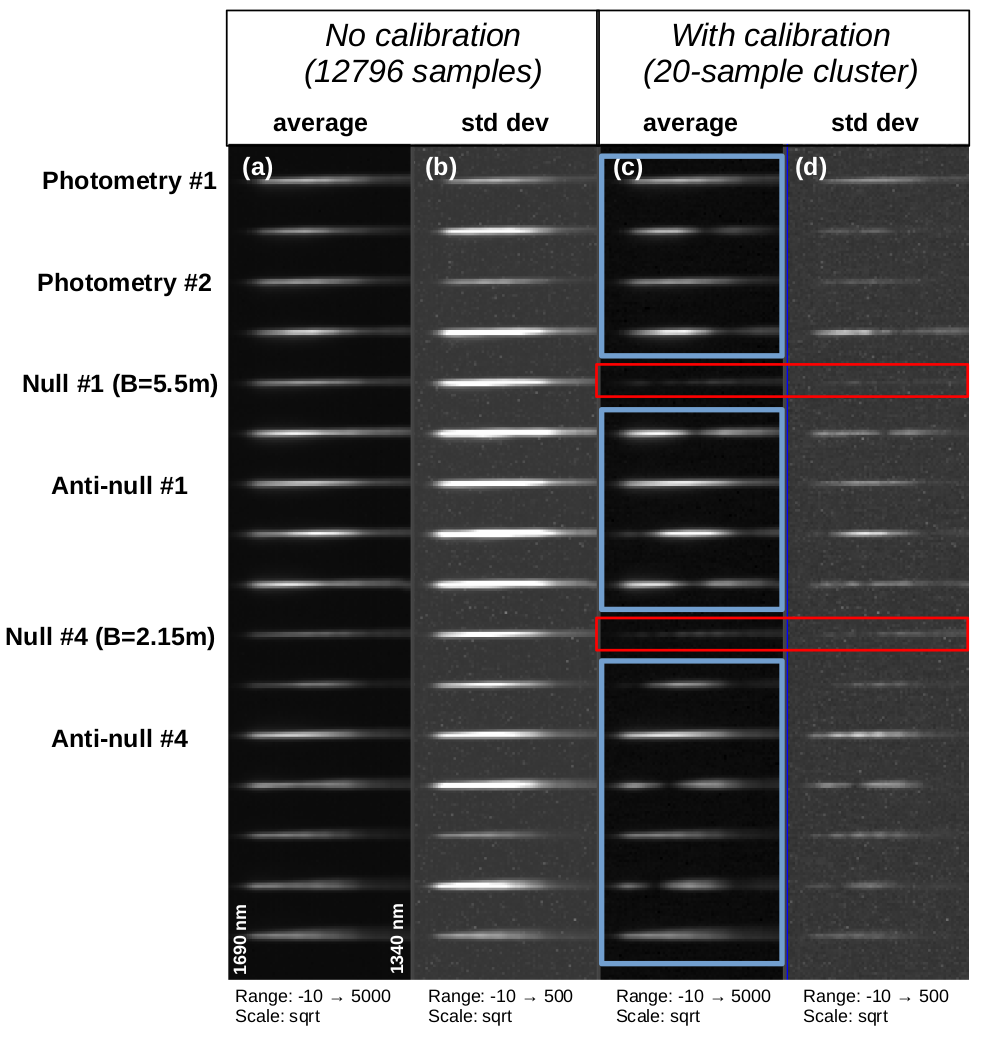}
    \includegraphics[width=8cm]{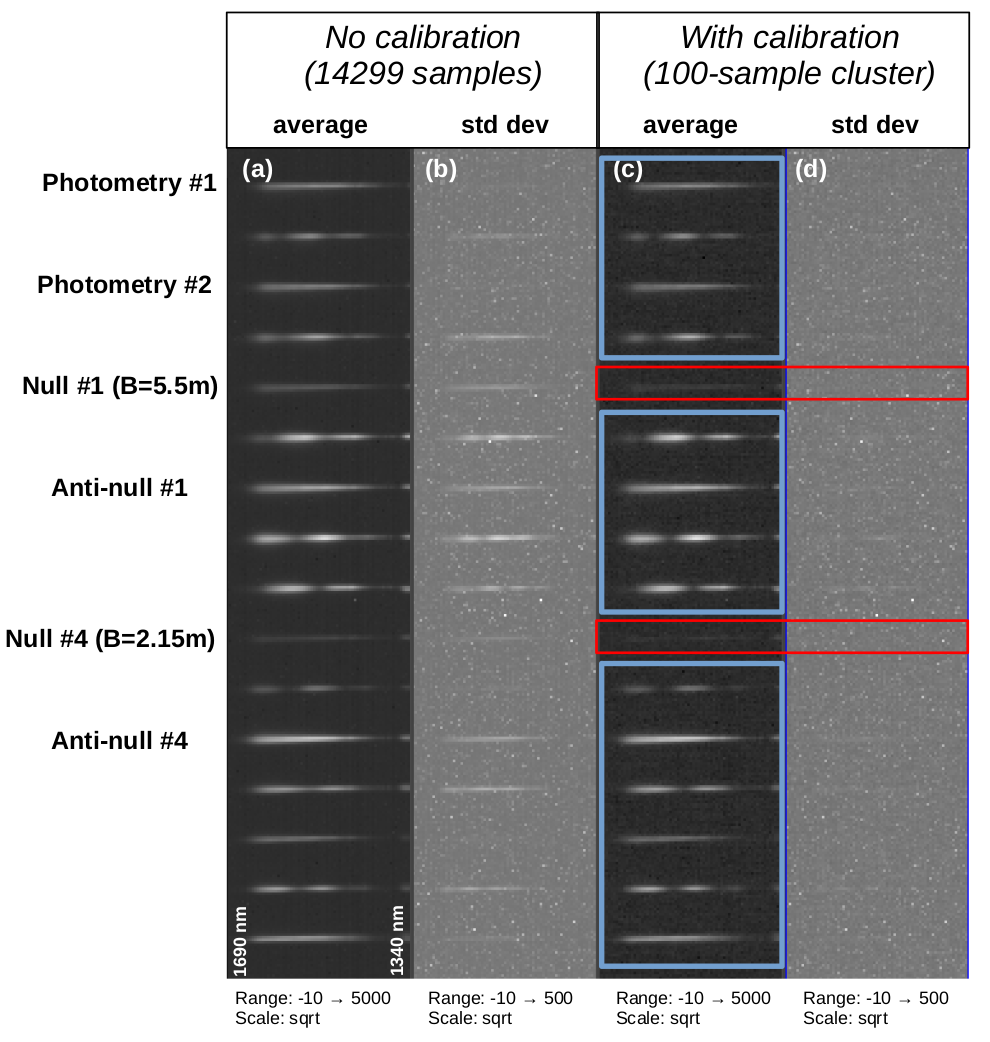}
    \vspace*{0.3cm}
    \caption{Null calibration with the GLINT photonic nuller: Laboratory (left) and on-sky (right) demonstration. Each of the 8 vertical panel shows the GLINT detector ouput, consisting of 16 horizontal spectra (one per output of the photonic chip) ranging from 1340 nm (right edge) to 1690 nm (left edge). Images are grouped by pair, with the average image on the left and standard deviation on the right. The leftmost pair is the average (a) and standard deviation (b) of the full set of 12,796 consecutive images. The average (c) and standard deviation (d) of one of the clusters that define the BF-to-DH is shown. Blue boxes indicate the BF used for selection, and the red boxes show the DH signals. The right side of the figure shows the same data for on-sky observations.}
    \label{fig:GLINTcalib}
\end{figure}

We validated the BF-to-DH calibration approach on the GLINT instrument\cite{Martinod2021NatCo} installed on the Subaru Telescope. Figure \ref{fig:GLINTcalib} shows laboratory (left) and on-sky (right) results. Wavefront errors were added with the system's deformable mirror in the lab experiment. All datasets were aquired at 1.4 kHz frame rate.

Both experiments demonstrate that the DH signal is well constrained by the BF state. In the on-sky result, the standard deviation of the DH signal (d) after calibration is below the readout and photon noise of the DH signal, indicating that the {\bf self-calibration is able to estimate the residual starlight to an accuracy below photon and readout noise}. Details of the experiments are provided in a previous publication \cite{2021SPIE11823E..18G}. The experiment validates uniqueness of the null solution for a given BF measurement, with no evidence of a measurement null space which would induce a variation in the null ouputs without a corresponding signature in the BF.

\section{Conclusion and Perspectives}

Self-calibration of high contrast imaging data using WFS telemetry is a promising solution to the current speckle noise limitation, potentially providing photon-noise limited detection limits on space and ground high contrast imaging systems. Preliminary on-sky experiments described in this paper are encouraging, and demonstrate that speckle noise is well constrained by WFS telemetry. The experiments were however limited in duration and did not validate that the WFS-to-PSF relationship is stable over long periods of time, which is critical to subtract speckle noise without removing planet signal. If the calibration holds, then the WFS-to-PSF mapping can be recorded on a calibration star (without a planet) and used to remove speckle noise on the science target. This demanding stability requirement will most likely be achieved by minimizing the time-variable non-common path aberrations (NCPA) between the input WFS and output science image.

Our findings suggest that future space-based high contrast imaging systems could be designed with less demanding telescope wavefront stability, as speckle noise can be accurately calibrated and removed from the science images. In this regime, speckle noise at or below the natural zodi+exozodi background at $\approx$ 3e9 contrast will not affect detetion limits, and stronger residual starlight can be compensated for by longer integration time to average photon noise. The coronagraph throughput and bandpass would be optimized for sensitivity in the photon-noise regime instead of reaching contrast levels below the exoplanet level.

For both space and ground systems, accurate self-calibration will require a stable or well-calibrated relationship between WFS and science data. The photonic nulling chip approach appears particularlty promising, as (1) the WFS and nulling functions are imprinted in the same physically small device, leaving little room for the WFS-to-DF relationship to change, (2) the WF state is discretized in a small number of coherent waveguides and (3) the output channels can easily be dispersed in wavelength.

We have relied on an explicit BF-to-DH mapping in this study. Our experiments built the mapping by clustering input samples and averaging the corresponding output samples. This allows for inspection of the output samples corresponding to the samples of the input cluster, so that the mapping uniqueness can be verified and quantified. This "lookup table" approach does not scale efficiently to high dimensions, for which more efficient reconstruction algorithms are needed. Neural network approaches have recently been shown to address this challenge \cite{Norris2022PL,Norris2022GLINT}, and perform well on the same photonic nulling data as presented in \S \ref{sec:GLINT}. Validation over larger datasets and longer time spans will be required to confirm algorithms are suitable for the high contrast imaging application discussed in this paper.

\acknowledgments

This work was supported by NASA grants \#80NSSC19K0336 and \#80NSSC19K0121. This work is based on data collected at Subaru Telescope, which is operated by the National Astronomical Observatory of Japan. The authors wish to recognize and acknowledge the very significant cultural role and reverence that the summit of Maunakea has always had within the Hawaiian community. We are most fortunate to have the opportunity to conduct observations from this mountain. The authors also wish to acknowledge the critical importance of the current and recent Subaru Observatory daycrew, technicians, telescope operators, computer support, and office staff employees.  Their expertise, ingenuity, and dedication is indispensable to the continued successful operation of these observatories. The development of SCExAO was supported by the Japan Society for the Promotion of Science (Grant-in-Aid for Research \#23340051, \#26220704, \#23103002, \#19H00703 \& \#19H00695), the Astrobiology Center of the National Institutes of Natural Sciences, Japan, the Mt Cuba Foundation and the director's contingency fund at Subaru Telescope. KA acknowledges support from the Heising-Simons foundation.

\bibliography{report} 
\bibliographystyle{spiebib} 

\end{document}